\documentstyle[astrobib,psfig]{mn-ab}
\def\cf{{\it cf.\ }}
\def\eg{{\it e.g.\ }}
\def\ie{{\it i.e.\ }}

\newcommand{\msun}{\mbox{$M_{\odot}$}}

\newcommand{\kmsmpc}{\mbox{km s$^{-1}$ Mpc$^{-1}$}}

\def\la{\mathrel{\hbox{\rlap{\hbox{\lower4pt\hbox{$\sim$}}}\hbox{$<$}}}}
\def\ga{\mathrel{\hbox{\rlap{\hbox{\lower4pt\hbox{$\sim$}}}\hbox{$>$}}}}

\def\pr{{\cal P}}
\title[How to Plant a Merger Tree]{How to Plant a Merger Tree}
\author[R.S. Somerville \& T.S. Kolatt]
       {Rachel S. Somerville$^{1,2}$ and Tsafrir S. Kolatt$^2$\\
        $^1$Racah Institute of Physics, The Hebrew University, Jerusalem\\
        $^2$Physics Department, University of California, Santa Cruz}

\begin{document}

\maketitle

\begin{abstract}
We investigate several approaches for constructing Monte Carlo realizations of
the merging history of virialized dark matter halos (``merger trees'') using
the extended Press-Schechter formalism. We describe several unsuccessful
methods in order to illustrate some of the difficult aspects of this
problem. We develop a practical method that leads to the reconstruction of the
mean quantities that can be derived from the Press-Schechter model. This method
is convenient, computationally efficient, and works for any power spectrum or
background cosmology. In addition, we investigate statistics that describe the
distribution of the number of progenitors and their masses as a function of
redshift.
\end{abstract}


\section{Introduction}
\label{sec:intro}
In the standard picture of modern structure formation, small-amplitude Gaussian
density fluctuations, which perhaps arose from quantum fluctuations and were
amplified by a period of rapid inflation, become more overdense with respect to
their surroundings as the universe expands. Eventually the self-gravity acting
on these regions becomes larger than the pressure of the expansion, and they
collapse to form bound, virialized structures. In hierarchical models, such as
the cold dark matter (CDM) family of models, the amplitude of the fluctuations
decreases with increasing scale. Thus small mass objects form first, and are
then incorporated into larger structures as time progresses. These dense,
gravitationally bound structures provide the environments where galaxies can
form. Hierarchical structure formation thus gives a natural explanation for the
very complex observed large scale structure of the Universe, \ie clusters,
superclusters, filaments, etc.

One way to study this process is with N-body simulations. However, numerical
simulations have familiar drawbacks. They are computationally expensive, so it
is difficult or impossible to explore a wide range of models or different
realizations of the same model. In addition memory and time limitations make it
impossible to attain the mass and force resolution required to simultaneously
study objects from dwarf galaxies ($\sim 10^9 \msun$) to clusters ($\sim
10^{15} \msun$). Semi-analytic methods are therefore an important alternative.

The model developed by \citeN{ps:74} provides a simple but relatively effective
framework for the description of the mass history of particles in a
hierarchical universe with Gaussian random phase initial perturbations. The
focus of the original Press-Schechter model was the derivation of the
multiplicity function of non-linear objects (``halos'') as a function of
redshift, \ie the ``mass function'' or number density of halos of a given mass
at a redshift $z$. This prediction has been tested quite extensively and found
to be in relatively good agreement with N-body simulations
\cite{efstathiou:88,gelb:94,lc:94,ma:96,gross:suites}. The Press-Schechter
theory was extended to give the \emph{conditional} probability that a particle
in a halo of mass $M_0$ at $z_0$ was in a halo of mass $M_1$ at an earlier
redshift $z_1$, leading to an expression for the conditional mass function
\cite{bower:91,bcek}. The extended Press-Schechter formalism can also be
manipulated to obtain expressions for halo survival times, formation times, and
merger rates \cite[hereafter LC93]{lc:93}, which have also been shown to agree
reasonably well with the results from N-body simulations \cite{lc:94}.

The computation of these mean quantities within the Press-Schechter model is
straightforward. However, for certain purposes one would like to go beyond
this. In particular, the semi-analytic approach to modeling galaxy formation
(\cf \citeNP{kwg}, \citeNP{cafnz}) attempts to describe the formation history
of galaxies and gas within dark matter halos, including simplified
hydrodynamics, star formation, supernova feedback, galaxy-galaxy merging, and
stellar population synthesis. These models rely on the construction of a
``merger tree'', which involves predicting the masses of progenitor halos and
the redshifts at which they merge to form larger halos. Galaxies initially form
in their own halo and are traced as they are incorporated into larger halos,
and eventually perhaps merge with other galaxies. A halo of a given mass may
have a variety of merging histories, and the properties of galaxies that form
within this halo presumably depend to some extent on the details of this
history.

Most of the previous work using semi-analytic models has focussed on
reproducing or predicting mean quantities and qualitative trends. However, as
observational data continues to improve, one would like to be able to
investigate whether the broader properties of the predicted \emph{distribution}
of model galaxies are consistent with the observations. For example, there has
already been some investigation of whether the scatter in the observed
Tully-Fisher relation \cite{eisenstein:96}, and in the color-magnitude and
line-strength velocity-dispersion (Mg-$\sigma$) relations \cite{kauffmann:96}
in merger models is consistent with observations. Before we can trust the
models for evaluating these kinds of questions, we must ensure that the merger
trees not only satisfy the mean properties readily predicted by the extended
Press-Schechter model, but ideally also the full distribution function. As we
shall see, accomplishing this goal is far from straightforward and has not been
thoroughly investigated in previous work on this subject.

In this paper, we discuss some of the practical and theoretical difficulties of
using the extended Press-Schechter model to create Monte Carlo merger histories
of dark matter halos. We mention some limitations of the previously proposed
methods for the construction of merger trees, and the motivation for developing
a new approach. We discuss several unsuccessful approaches, in an attempt to
clarify some aspects of this problem as well as to prevent others from
following the same dead ends. In addition, the formalism we present may be
useful in future work on this subject, as the final approach that we present
seems to be effective and convenient, but it is still not rigorous. A primary
motivation for embarking on this project was to develop a method that
reproduces the full joint probability distribution, rather than just the
mean. Our method is not guaranteed to do so, and further investigation of this
question requires a comparison with N-body simulations. This will be presented
in a companion paper \cite{paperII}.

In \S\ref{sec:ps}, we give a brief introduction to the Press-Schechter
formalism. In \S\ref{sec:previous} we summarize some of the previous methods
for creating merger trees. In \S\ref{sec:difficulties}, we describe the source
of some difficulties one encounters in attempting to use the extended
Press-Schechter formalism to construct merger trees. We develop a simple model
of the joint probability distribution in \S\ref{sec:gestalt}, and attempt to
use it to construct merger trees. In \S\ref{sec:practical}, we present a
completely different approach which eventually leads to our successful method,
described in \S\ref{sec:final}. In \S\ref{sec:distribution}, we investigate
the distribution of progenitor number and mass given by our successful method.
We summarize and conclude in \S\ref{sec:sum}.

Readers who are only interested in the successful method may skip to
\S\ref{sec:practical}. 

\section{The Press-Schechter Formalism}
\label{sec:ps}
The Press-Schechter model \cite{ps:74} is based on a combination of linear
growth theory, spherical collapse theory, and the properties of Gaussian random
fields. Suppose that we have smoothed the initial density distribution on a
scale $R$ using some spherically symmetric window function $W_{M}(r)$, where
$M(R)$ is the average mass contained within the window function. There are
various possible choices for the form of the window function (\cf
\citeNP{lc:93}), and the relation between $M$ and $R$ will clearly depend on
this choice. We use a real-space top hat window function, $W_M(r) =
\Theta(R-r)(4\pi R^3/3)^{-1}$, where $\Theta$ is the Heaviside step
function. In this case $M = 4\pi \rho_0 R^3/3$, where $\rho_0$ is the mean mass
density of the universe. The mass variance $S(M) \equiv \sigma^2(M)$ may be
calculated from
\begin{equation}
\label{eqn:massvariance}
\sigma^2(M) = \frac{1}{2\pi^2} \int P(k) W^2(kR) k^2 {\rm d}k \, ,
\end{equation}
where $P(k)$ is the mass power spectrum, and $W(kR)$ is the Fourier
transform of the real space top-hat:
\begin{equation}
W(kR) = \frac{3[\sin(kR) - kR\cos(kR)]}{(kR)^3}.
\end{equation}

The ``excursion set'' derivation due to \citeN{bcek} leads naturally to the
extended Press-Schechter formalism that we will use extensively in this
paper. The smoothed field $\delta(M)$ is a Gaussian random variable with zero
mean and variance $S$. The value of $\delta$ executes a random walk as the
smoothing scale is changed. Adopting an ansatz similar to that of the original
Press-Schechter model, we associate the fraction of matter in collapsed objects
in the mass interval $M, M+{\rm d}M$ at time $t$ with the fraction of
trajectories that make their \emph{first upcrossing} through the threshold
$\omega \equiv \delta_c(t)$ in the interval $S, S+{\rm d}S$. This may be
translated to a mass interval through equation (\ref{eqn:massvariance}). The
halo multiplicity function (here in the notation of LC93) is then:
\begin{equation}
\label{eqn:ps}
f(S, \omega) {\rm d}S = \frac{1}{\sqrt{2\pi}} \frac{\omega}{S^{3/2}} 
\exp{\left[-\frac{\omega^2}{2S}\right]} {\rm d}S \, .
\end{equation}

The \emph{conditional} mass function, the fraction of the trajectories in halos
with mass $M_1$ at $z_1$ that are in halos with mass $M_0$ at $z_0$ ($M_1 <
M_0$, $z_0 < z_1$) is 
\begin{eqnarray}
\label{eqn:flc}
\lefteqn{f(S_1, \omega_1 \mid S_0, \omega_0) {\rm d}S_1 =} 
\nonumber \hspace{1truecm}\\
 & &  \frac{1}{\sqrt{2\pi}} \frac{(\omega_1-\omega_0)}{(S_1-S_0)^{3/2}}
\exp{\left[-\frac{(\omega_1-\omega_0)^2}{2(S_1-S_0)}\right]} {\rm d}S_1
\,.
\end{eqnarray}
The probability that a halo of mass $M_0$ at redshift $z_0$ had a
progenitor in the mass range $(M_1, M_1+{\rm d}M_1)$ is given by (LC93):
\begin{eqnarray}
\label{eqn:Nlc}
\lefteqn{\frac{{\rm d}P}{{\rm d}M_1}(M_1, z_1 \mid M_0, z_0) {\rm d}M_1 =}
\nonumber \hspace{2.5truecm}\\
& & \frac{M_0}{M_1} f(S_1, \omega_1 \mid S_0, \omega_0) 
\left| \frac{{\rm d}S}{{\rm d}M}\right| {\rm d}M_1\, ,
\end{eqnarray}
where the factor $M_0/M_1$ converts the counting from mass weighting to number
weighting. 

All of the results presented in this paper have been calculated for an
$\Omega=1$ universe with $H_0=50\; \kmsmpc$. The power spectrum is obtained
from the fitting formula of \citeN{bbks} with $\Gamma=0.21$ and $\sigma_8 =
0.6$. This is the $\tau$CDM model of \citeN{ebw:92}, and has been chosen
because the slope of the power spectrum on galaxy scales is consistent with
observations. However, our results are equally valid for any assumed power
spectrum or cosmology. The one exception that we know of is the case of a
universe with a ``hot'' dark matter component, such as a massive neutrino (CHDM
or MDM type models). The standard extended Press-Schechter formalism does not
properly treat the evolution due to the changing free-streaming length of the
neutrino in such models. All of the expressions given are valid for a general
cosmology unless otherwise noted.

\section{Previous Methods}
\label{sec:previous}
The development of techniques for constructing Monte Carlo realizations of the
merging history of dark matter halos \cite{kw,block1,block2,lc:93} using the
extended Press-Schechter formalism has allowed a great deal of progress to be
made in the use of semi-analytic methods for studying galaxy formation and
evolution. \citeN{kw} developed a method for constructing merger trees which
addresses the problem of simultaneously reproducing the average number of halos
given by Eqn.~\ref{eqn:Nlc} and imposing the constraint that the mass of a
halo be equal to the sum of the masses of its progenitors at every stage. To do
this they impose a grid in mass and redshift. For the first step in redshift,
they then create a list of halos, where the number of halos with mass $M_i$ is
given by $N(M_i) = N_{\rm ens}\: {\rm d}N/{\rm d}M(M_i)\: \Delta M_i$, rounded to
the nearest integer. $N_{\rm ens}$ is some large number of ensembles, typically
$N_{\rm ens} \sim 100$. The progenitors are randomly assigned to ensembles,
starting with the largest and working in order of decreasing mass. The
probability of assignment is proportional to the amount of remaining free mass,
with the constraint that the total mass of the progenitors cannot exceed the
mass of the parent\footnote{Note that in our terminology, the ``parent'' is
younger than its ``progenitors'', because we always work backwards in
time.}. This process is repeated for all the steps in the redshift grid.

This algorithm is guaranteed to exactly reproduce the mean number of halos of
each mass at each redshift for the set of ensembles. However, it is possible to
encounter a situation in which the next halo does not fit into any of the
ensembles. In order to get around this problem, mass conservation is only
enforced in an approximate way (G. Kauffmann, private communication). There are
also some practical drawbacks to this method. It is necessary to generate a
large number of ensembles and store them, which is somewhat inconvenient. An
arbitrary grid in halo mass and redshift must be imposed. Also, because the
function ${\rm d}N/{\rm d}M$ is very sharply peaked around $M_0$ for small
$M_0$ or small redshift intervals $\Delta z$, the algorithm as described in
\citeN{kw} is sensitive to the binning used and is prone to numerical
problems. The algorithm breaks down for certain choices of power
spectrum. Finally, although the mean of the distribution is reproduced by
construction, the partitioning of the halos into individual ensembles is ad-hoc
and may or may not reproduce the higher moments of the distribution.

A different approach, referred to as the ``Block model'', has been proposed by
\citeN{block1} and \citeN{block2}. A major drawback with this approach is that
the halo masses always grow in discrete steps of factors of two. This is
problematic for the purpose of semi-analytic galaxy formation modeling, in
which one would like to follow individual galaxies with fairly fine time
resolution. LC93 propose a generalization of the block model which removes this
condition, but we show in Section~\ref{sec:bintree} that this method produces
halo mass distributions that are severely discrepant with the Press-Schechter
model.

\section{Difficulties of Trees}
\label{sec:difficulties}
The first choice we must make if we want to construct a merger tree is whether
to start from high redshift and merge together small clumps until the desired
redshift is reached (as in an N-body simulation or, presumably, in the real
universe), or, to start from the present day and work backwards in time,
``disintegrating'' the halos into their progenitors like a film run
backwards. The extended Press-Schechter formalism provides expressions
applicable to both situations. An important consideration is that we would like
to eventually anchor our approach using the observed $z=0$ properties of
galaxies, which are presumably the most secure. In addition, because we lack
any information about spatial correlations, if we go forward in time, we do not
know which small clumps to combine with which. It may be possible to get around
this problem somehow, but for now we pursue the ``disintegration'' approach, in
which we postulate the existence of a ``parent'' halo of a given mass $M_0$ at
redshift $z_0$ and break it into its ``progenitors'' working backwards in time.

It might appear a simple matter to construct a merger tree by simply picking
the masses of the progenitors of our parent halo at some earlier redshift $z_1$
from the expression for ${\rm d}P/{\rm d}M(M_1, z_1 \vert M_0, z_0)$ given by
Eqn.~\ref{eqn:Nlc} above, then repeating this process starting from each
progenitor in turn for the next step back in time. Two difficulties immediately
arise in implementing this approach. First, from inspection of
Eqn.~\ref{eqn:Nlc}, the {\it number} of halos clearly diverges as the mass
goes to zero. However, note that the {\it mass} contained in small halos
(Eqn.~\ref{eqn:flc}) does not diverge as $M \rightarrow 0$. In order to
pick masses from the number weighted probability function numerically, it is
necessary to introduce a cutoff mass, or effective mass resolution, $M_l$.

The second problem is that the progenitor masses must simultaneously be drawn
from the distribution ${\rm d}P/{\rm d}M(M)$ \emph{and} add up to the mass of
the parent, $M_0$. The problem is that ${\rm d}P/{\rm d}M$ is just the
\emph{average} number of halos that one can make out of the mass $M_0 f(M) \,
{\rm d}M$. What we really want is the \emph{joint} probability function for the
set of progenitors $\{M_1, \cdots, M_n\}$, ${\rm d}P_n/{\rm d}M(\{M_1, \cdots,
M_n\}, z_1 \mid M_0, z_0)$ with any value of $n<M_0/M_l$
\footnote{From now on we will drop the differential notation ${\rm d}P/{\rm
d}M$ and refer to the probability given by Eqn.~\ref{eqn:Nlc} as simply
$P(M)$. We will also frequently drop the explicit dependence on the redshift
and the parent mass $M_0$ where this is unambiguous.}. An obvious problem with
the use of the single halo probability rather than the joint probability is
that there is no guarantee that we will not at some stage pick a progenitor
that does not ``fit'' in the halo: \ie $M > M_0 - \sum_{i} M_i$ where $M_i$ are
the masses of all the previously picked progenitors. In addition, since the
expression $P(M)$ only gives the probability that there was \emph{a} progenitor
of mass $M$ at an earlier time, we do not know {\it a priori} how many
progenitors were present at the redshift $z_1$.

Having noted these points, we can write down some basic requirements for our
merger tree construction algorithm:
\begin{enumerate}
\item
The procedure must account for the mass contained in halos below the mass
resolution $M_l$, which we will refer to as the ``accreted mass''. However the
results must be independent of the value of $M_l$.
\item
The procedure should treat all progenitors equally, independently of the
sequence in which they are chosen. 
\item
The procedure must simultaneously reproduce the distribution of the
number of progenitors and their masses while conserving mass.
\item
The algorithm should be numerically robust and must be possible to
implement in a computationally efficient and convenient way.
\end{enumerate}

\begin{figure}
\centerline{\psfig{file=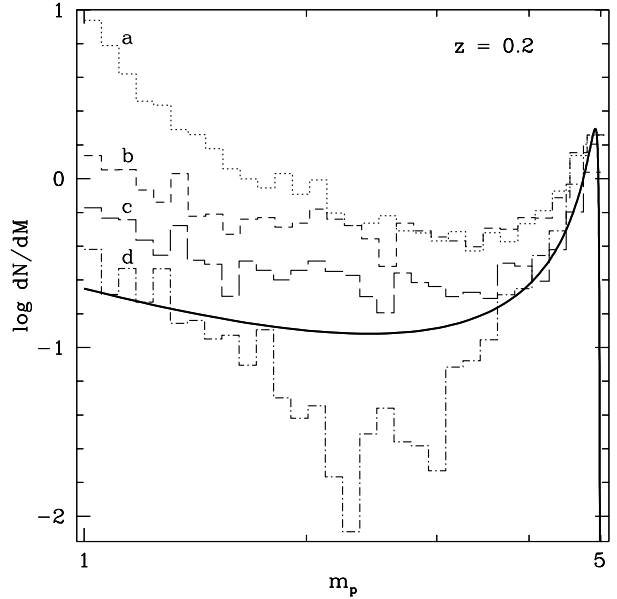,width=\columnwidth}}
\caption{The number of progenitors with mass $m_p$ (conditional mass function)
for a single step in redshift ($z_0=0$ to $z_1=0.2$). The bold solid line is
the prediction of the extended Press-Schechter theory
(Eqn.~\protect\ref{eqn:Nlc}). The histograms were obtained by picking masses
from the distribution $P(M,z_1 \mid M_0, z_0)$ until the parent mass $M_0$ was
exhausted. The dotted histogram (a) shows the results using the Na\"{\i}ve
Method with Cutoff, and the short-dashed histogram (b) shows the results of
using the Na\"{\i}ve Method with Truncation (see text). The long-dashed
histogram (c) uses Accretion Model Method 1, and the dot-dashed histogram (d)
uses Accretion Model Method 2 (see text). All masses are in units of the
minimum progenitor mass $M_l$.}
\label{fig:naive}
\end{figure}

We now demonstrate some problems that arise in several seemingly
straightforward approaches to building the trees. The bold solid line in
Figure~\ref{fig:naive} shows the prediction of the extended Press-Schechter
model for the quantity ${\rm d}N/{\rm d}M$, the number of progenitors with mass
$M$ for a parent halo with $M_0 = 5\,M_l$ after a single step in redshift from
$z_0=0$ to $z_1=0.2$ (also called the conditional mass function). This is the
quantity that a successful merging tree method must reproduce. In this figure
and hereafter unless otherwise noted, all masses are given in units of the mass
resolution $M_l$. Here we have used $M_l = 1.0\times 10^{10} \msun$, but the
results are independent of this value.

In our first attempted method, the progenitor masses are chosen from
Eqn.~\ref{eqn:Nlc} until the mass reservoir $M_0$ is exhausted. The
probability is set to zero for $M < M_l$. We have tried two ways of addressing
the problem of mass ``overflow'' described above. One approach is to choose
progenitors until the total mass exceeds $M_0$, and then to truncate the mass
of the last progenitor. We will refer to this as the ``Na\"{\i}ve Method with
Truncation''. Another approach is to impose an upper mass cutoff, so that the
probability is effectively set to zero for any values of $M$ that exceed the
available mass. In effect we then choose the $i$-th progenitor from the
modified distribution
\begin{equation}
\label{eqn:Nlc2}
P_i(M_i) = P(M_i) \Theta(M_i-M_l) \Theta(M_0-\sum_{j=1}^{i} M_j) \, ,
\end{equation}
where $P(M)$ is the original probability function from Eqn.~\ref{eqn:Nlc}
renormalized for its new range. No attempt is made to compensate for the
contribution of masses below $M_l$. We refer to this as the ``Na\"{\i}ve Method
with Cutoff''. The results for the quantity ${\rm d}N/{\rm d}M$ (averaged over
many ensembles) is shown by the histograms marked a (Cutoff) and b (Truncation)
in Figure~\ref{fig:naive}.

Both procedures clearly fail to correctly reproduce the conditional mass
function predicted by the extended Press-Schechter model
(Eqn.~\ref{eqn:Nlc}). Although the introduction of the upper mass cutoff to
prevent choosing masses that are too large to fit in the current ensemble is
somewhat more elegant than the brute force truncation, we see that it produces
a shift towards smaller masses which leads to large excess of small mass
halos. It should be noted that this figure shows only one step in redshift. The
excess multiplies with each subsequent step in redshift, so that even a
relatively small discrepancy quickly becomes very serious. It should also be
noted that these problems are most pronounced when $M_0 \approx M_l$ (as in the
case shown in the figure). If $M_0 \gg M_l$, the disagreement is not as bad.

\section{The Accretion Model}
\label{sec:gestalt}
We must introduce a self-consistent way of treating progenitors above and below
the cutoff mass, $M_l$. It should be noted that although the contribution of
masses $M<M_l$ may be negligible for halos with $M_0 \gg M_l$, as the mass of
the parent approaches the resolution limit it necessarily becomes a significant
fraction of the total progenitor mass. Because every halo, regardless of its
size, must be broken down into smaller and smaller pieces until all of the
pieces fall below the mass resolution (this is what makes the process finite),
the correct treatment of small halos is crucial for reconstructing the
formation history of halos of all masses out to arbitrarily high redshift.

The basic idea behind this approach is never to treat any mass below $M_l$ in
terms of progenitor number, but rather to find a complimentary description for
it as accreted diffuse matter. We now introduce an arbitrary distinction in
terminology to reflect this division. Let {\it progenitors} by definition have
mass greater than the fixed mass resolution $M_l$. The aggregate contribution
of all halos with $M < M_l$ will be referred to as {\it accreted mass}. A fully
rigorous procedure should use the {\it joint} probability for progenitor number
(not mass) above $M_l$, and accreted mass (not number of halos) below this mass
scale.

We can now define a few more useful quantities. Given the mass of the parent
halo $M_0$ and the redshift step $z_0 \rightarrow z_1$, the average number of
progenitors (recalling our definition above), is:
\begin{equation}
\label{eq:avg_n_p}
\bar{N} \equiv
\langle N_p (M \,\vert M_0) \rangle = \int_{M_l}^{M_0} \, {\rm d}M\,
\frac{M_0}{M}\, P(M,z_1 \vert M_0, z_0) 
\,.
\end{equation}
We can also calculate the average fraction of $M_0$ that dwelt
in the form of progenitor halos of mass $M>M_l$:
\begin{equation}
\label{eqn:fproj}
\bar{f}_p =\int_{M_l}^\infty{\rm d}M\, P(M,z_1 \vert M_0,z_0)\,,
\end{equation}
and the complimentary quantity for the average fraction of $M_0$ that
came from ``accreted'' mass, $\bar{f}_{acc} =1-\bar{f}_p$.

Before we proceed, we would like to warn the reader that the contents of the
remainder of this section are rather detailed and probably only of interest to
the specialist. The formalism developed in the rest of this section is not used
directly in the successful method that we will eventually derive. We urge the
impatient reader to skip directly to \S\ref{sec:practical}.

From the above predictions we can try to evaluate what went wrong with our
previous procedures (the ``Na\"{\i}ve'' methods). The predicted average number
of progenitors for the case considered in Figure~\ref{fig:naive} ($M_0 =
5\,M_l$, $z_0=0$, $z_1=0.2$) is $\bar{N}=1.14$. The actual average number for
one hundred Monte Carlo realizations using the Na\"{\i}ve Method with
Truncation is $\bar{N}=2.1$, and for the Na\"{\i}ve Method with Cutoff it is
$\bar{N}=2.3$.  Thus we see that the mean of the distribution is shifted
towards larger numbers of small mass halos. This motivates our goal in this
section, which is to find the probability function $\pr_N$ of having $N$
progenitors at redshift $z_1$ given the mass $M_0$ at a later redshift $z_0$,
with the imposed cutoff of $M_l$.

Given $M_0$, $z_0$, and $z_1$, let $M_{1,p}$ be the mass of a progenitor, where
by definition $M_{1,p}>M_l$. However, during the time interval $\Delta z \equiv
z_1-z_0$, this progenitor accretes a mass $M_{1,\rm acc}$ due to merging with
small halos of mass $M<M_l$ which are not counted as progenitors. Therefore its
effective contribution to $M_0$ is $M_1=M_{1,p}+M_{1, \rm acc}$. We now define
a modified probability function
\begin{equation}
\label{eq:p_mod_1}
\tilde{P} (M_1\vert M_0)=\int_{M_l}^{M_1} {\rm d}M_{1,p}\,
P(M_{1,p}\vert M_0) P_{\rm acc}(M_{1, \rm acc} \vert M_{1,p}).
\end{equation}
The weighting function $P_{\rm acc}$ is proportional to the probability for
$M_{1,p}$ to accrete a mass $M_{1,\rm acc}$ during the specified redshift
interval. This probability {\it is not} simply $P(M_{1,\rm acc},z_1 \vert M_0,
z_0)$, because $M_{\rm acc}$ will in general be comprised of many small
halos. It should be noted that for the same reason, $M_{\rm acc}$ is not
necessarily smaller than $M_l$. We return to the determination of the function
$P_{\rm acc}$ in a moment.

The probability for having one and only one progenitor (with additional
accreted mass) given a halo of mass $M_0$ is 
\begin{equation}
\pr_1 = \tilde{P} (M_0 \vert M_0)\,.
\end{equation}
The probability for exactly two progenitors is
\begin{equation}
\pr_2  =  \int_{M_l}^{M_0-M_l}\, {\rm d}M_1\,\tilde{P}_{2} (M_1\vert M_0)\, ,
\end{equation}
where
\begin{equation}
\tilde{P}_{2} (M_1 \vert M_0) \equiv 
\tilde{P}_{1} (M_1 \vert M_0) \tilde{P}_{1} (M_0-M_1 \vert M_0)\,.
\end{equation}
The generalization to $N$ progenitors $\pr_N$ is obtained recursively
via
\begin{eqnarray}
\pr_N & = & \int_{M_l}^{M_0-(N-1)M_l}{\rm d}M_1 \tilde{P}_{N} (M_1,...,M_N
\vert M_0) \\
&=& \int_{M_l}^{M_0-(N-1)M_l}{\rm d}M_1\,\tilde{P}_{1} (M_1 \vert M_0)
\nonumber\\
&\cdots & \int_{M_l}^{M_0-\sum\limits_{i=1}^{j-1}M_i-(N-j-1)M_l}
{\rm d}M_j \tilde{P}_{1} (M_j \vert M_0)  \nonumber \\
& \cdots & 
\int_{M_l}^{M_0-  \sum\limits_{i=1}^{N-2}(M_i) -(N-2)M_l}{\rm d}
M_{N-1} \nonumber \\
&& \; \times \tilde{P}_{1} (M_{N-1} \vert M_0)\,
\tilde{P}_{1} (M_0-\small{\sum\limits _{i=1}^{N-1}(M_i)} \vert \, M_0) \,.
\label{eq:prn}
\end{eqnarray}
The probability of having no progenitors of mass bigger than $M_l$, $\pr_0$, is
evaluated at the end by the requirement of
\begin{equation}
\sum \limits_{i=0}^N\pr_i = 1 \,,
\end{equation}
where $N$ is sufficiently large that $\pr_N \rightarrow 0$. The average number
of progenitors must satisfy
\begin{equation}
\langle N_p \rangle = \sum \limits_{n=0}^N n\pr_n  \,,
\end{equation}
and may be compared with the independent prediction of Eqn.~\ref{eq:avg_n_p}. 

\subsection{The Accretion Probability}
\label{sec:accretion}
The accretion weighting function $P_{\rm acc}(M_{\rm acc} \mid M_p; z_1,z_0)$
is an important missing ingredient in these expressions. It is proportional to
the probability for a progenitor with mass $M_p$ to accrete a mass $M_{\rm
acc}$ during the redshift interval $\Delta z = z_1-z_0$. It should reflect our
expectation that it is very unlikely for a small mass halo to accrete a very
large amount of mass, as this would require the simultaneous merging of a very
large number of halos with $M<M_l$. Similarly, a large halo will be unlikely to
accrete a very small amount of mass, because its cross section for merging is
large. In this section we incorporate these qualitative expectations into a
reasonable guess for the accretion probability function $P_{\rm acc}$.

Consider a halo with a mass $M_1$ at a redshift $z_1$. From the
spherical collapse model (\eg \citeNP{wf}), we expect the virial
mass to increase due to the infall of previously uncollapsed
material. The mass at a later time corresponding to a redshift
$z_2 < z_1$ is
\begin{equation}
\label{eq:acc_avg}
M_{2} = M_1 + \int_{z_1}^{z_2}{\rm d} z\,\frac{{\rm d}M}
{{\rm d}t}(M_1)\, \frac{{\rm d}t}{{\rm d}z} \,,
\end{equation}
where the accretion rate from the spherical infall model is:
\begin{equation}
\frac{ {\rm d}M}{{\rm d}t }(M) = \frac{V_c^3}{2\pi G}  \,,
\end{equation}
where $V_c$ is the circular velocity of the halo (the last formula is strictly
true only in a universe with no cosmological constant, but is a good
approximation even if $\Lambda \neq 0$). This change in mass includes mergers
with halos of all masses. We still need to estimate how much of the mass change
$\Delta M = M_2-M_1$ is due to mergers with halos with mass less than the
resolution limit, \ie ``accretion''. To do this we use the expression for the
mean fraction of accreted mass ($\bar{f}_{\rm acc} \equiv 1-\bar{f}_{p}$,
starting from the mass $M_2$ and going back in time from $z_2$ to $z_1$. In
this way we estimate the \emph{average} mass accreted by $M_1$ to be
$\bar{M}_{\rm acc} = \bar{f}_{\rm acc}(M_2) \, M_2$.

For higher moments of the distribution function $P_{\rm acc}(M_{\rm acc}\vert
M_p;\,z_1,z_0)$ we shall assume that the accretion is mainly due to the infall
of blobs (\ie halos of mass $<M_l$) with typical mass $M_b \ll M_l$. For the
average accretion of Eq. \ref{eq:acc_avg} we expect
\begin{equation}
N_b \simeq \frac{\bar{M}_{\rm acc}}{M_b} \,.
\end{equation}
For any total accreted mass regardless of the value of $M_b$, the
number of blobs is proportional to the average accreted mass. If the
number of the blobs is Poisson distributed, then the second moment of
the accreted mass distribution function is proportional to the total
mass accreted. In the limit $N_b \gg 1$ the distribution should
approach a Gaussian distribution. We have already argued that the mean
of this distribution should be $\bar{M}_{\rm acc}$. We can make a rough
guess for the width of the distribution, $\sigma^2_{\rm acc} = \beta (\Delta
M-\bar{M}_{\rm acc})$ where $\Delta M$ represents the mass change
predicted by the spherical infall model. This should be an upper limit
on the accreted mass. Due to the uncertainties involved in the
derivation of this expression, the parameter $\beta$ is left free and
can be tuned as needed (we used $\beta=2$ for the results presented
here). We now have a reasonable guess for the functional form of the
accretion probability function for a progenitor of mass $M_p$ over the
redshift interval $z_1$ to $z_0$:
\begin{equation}
\label{eq:p_m_acc}
P_{\rm acc}(M_{\rm acc}\vert M_p;\, z_1,z_0) \propto \frac{1}{\sigma_{\rm
acc}^{1/2}}\, \exp \left[ - \frac{(M_{\rm acc} - \bar{M}_{\rm
acc})^2}{2\sigma^2_{\rm acc} } \right] .
\end{equation}
Although this expression is admittedly ad hoc, one can see that it
contains the correct qualitative behavior. The mean accreted mass
increases with the progenitor mass and with $\Delta z$ as expected.

\subsection{Merger Trees with the Accretion Model}
\label{sec:acctree}
We can now imagine a new approach for constructing the merger trees which
addresses the two main sources of the problems in the previous approach: the
failure to account for accreted mass and the incorrect distribution of the
number of progenitors. Given the probability function $\pr_N$ for each $M_0$
and time-step, we pick the number of progenitors from this distribution. We
assign a mass to each of these progenitors from the distribution $P_i(M)$
(Eqn.~\ref{eqn:Nlc2}) as before. The $\Theta$ function prevents us from
choosing a mass larger than the available mass at any stage. The accreted mass
is automatically obtained from the residue of this procedure. The results of
this algorithm (which we will call Accretion Model Method 1) are shown by the
histogram c in Figure~\ref{fig:naive}. We see that the results
have improved dramatically from the Na\"{\i}ve Method with Truncation where the
upper mass cutoff was also used but the number of progenitors was not
specified. Also the average number of progenitors is now $\bar{N}=1.08$, in
much better agreement with the expected value of $\bar{N}=1.14$. However there
is still an inconsistency in this procedure which leads to the remaining
discrepancy. We have still assigned the mass to the progenitors based on the
\emph{single} halo probability function $P(M)$, when, as we have argued before,
we really should have used the joint probability function $P_n(\{M_1, \cdots,
M_n\})$. This can in principle be calculated using the same approach that we
used to obtain the $\pr_N$ function. For example the joint probability for
$N=2$, taking into account the accretion weighting, is just the integrand of
the expression for $\pr_2$:
\begin{eqnarray}
\label{eqn:jointprob}
\lefteqn{P_2(\{M_1,M_2\}) = P(M_1)P(M_2)\, \Theta(M_0-M_1-M_2)} 
\nonumber \hspace{2truecm}\\
&& \times \int_{0}^{M_0-M_1-M_2} 
{\rm d}M_{\rm acc}\, P_{\rm acc}(M_{\rm acc} \mid M_1)\, 
\nonumber\hspace{2.5truecm} \\
&& \times P_{\rm acc}(M_0-M_1-M_2-M_{\rm acc} \mid M_2) \, .
\end{eqnarray}
This expression may be generalized to the $N$-halo joint probability
as before. 

We now pick the number of progenitors from $\pr_N$ and assign the masses from
the joint probability function Eqn.~\ref{eqn:jointprob} (Accretion Model
Method 2). This approach goes a long way towards curing the problems we have
noted, as we see from histogram d in Figure~\ref{fig:naive}. However the shape
of the mass function is not quite right. We attribute this to the inconsistency
introduced by our ad-hoc accretion probability weighting. We find that changing
the form of this function significantly affects the results obtained for the
mass function. We obtain better results for a lognormal distribution than for
the Gaussian distribution used here. If we could somehow obtain some external
information on the form of the accretion weighting, for example from N-body
simulations, it might be possible to produce a successfully working method.
However, this scheme is also rather cumbersome and computationally expensive.
The calculation of $\pr_N$ involves the computation of $(N-1)!$ integrals, and
must be repeated for every parent halo mass, $M_0$ and redshift interval
$\Delta z$. The joint probability function for the $i$-th progenitor will
depend on the masses of the previously chosen progenitors and thus must be
recalculated at each stage. This quickly becomes prohibitively computationally
expensive when large numbers of progenitors are allowed.

We can address the second problem by reducing the time-step or redshift
interval $\Delta z$. As $\Delta z$ is decreased, the form of $\pr_N$ steepens
and becomes peaked at smaller $N$. For a small enough choice of $\Delta z$,
$\pr _{N \ge 3} \rightarrow 0$. We will refer to this condition as the
two-progenitor limit. The idea is not to allow any processes that involve more
than a single bifurcation. This demand allows one to calculate only three
functions for each time step ($\pr_0$, $\pr_1$, \& $\pr_2$), by using the
analytic model of Eqn.~\ref{eq:prn}. At each stage, the time-step will now
depend on the parent mass $M_0$. The larger the halo, the smaller the time-step
necessary to satisfy this condition.

Although going to the two-progenitor limit might make the procedure
computationally feasible, for the moment our lack of knowledge about the
accretion probability weighting, and the sensitivity of the results to this
function, lead us to abandon this approach. Perhaps the formalism we have
developed here, and the simple approach we have presented for modeling the
joint probability function, can be refined in the future. However, we do not
pursue it in this paper.

\section{A Practical Solution}
\label{sec:practical}

\subsection{Binary Merger Trees Without Accretion}
\label{sec:bintree}
In the absence of external information about the behavior of the accreted mass
component, we are forced to treat it within our Monte Carlo procedure. As we
have discussed, we do not want to use the number-weighted probability for
masses below $M_l$ because of the divergence of this expression at small
masses. However, the \emph{mass}-weighted probability, Eqn.~\ref{eqn:flc}
does not diverge. If we pick a mass $M_1$ from the mass-weighted expression
$f(M_1, z_1 \mid M_0, z_0)$, this is equivalent to discovering that a single
trajectory, or particle, from the parent halo $M_0$ was in a halo with mass
$M_1$ at $z_1$. Once again, this is the single trajectory probability and if we
continue to select masses they will not in general fit together in any sensible
combination that can lead to $M_0$. We attempt to evade this problem by
choosing a very small time step and so going to the two progenitor limit, as
before. It is convenient to use $\omega \equiv \delta_c(z) = \delta_{c,0}/D(z)$
as our time variable and $S(M) \equiv \sigma^2(M)$ as our mass variable as in
LC93. These can be translated back to redshift and mass by inversion of the
appropriate expressions. The probability for a step $\Delta S$ in a time step
$\Delta \omega$ is (LC93; Eqn. 2.29)
\begin{equation}
\label{eqn:prob}
P(\Delta S, \Delta \omega) {\rm d}\Delta S 
= \frac{1}{\sqrt{2\pi}} \frac{\Delta \omega}{(\Delta S)^{3/2}} 
\exp \left[ -\frac{(\Delta \omega)^2}{2\Delta S} \right]
{\rm d}\Delta S
\end{equation}
If we make a change in variables $x \equiv \Delta \omega/(2 \sqrt{\Delta S})$
this becomes a Gaussian distribution in $x$ with zero mean and unit variance. 
We can see from this expression that if we choose the timestep such that 
\begin{equation}
\label{eqn:timestep}
\Delta \omega \la \sqrt{\frac{{\rm d}S}{{\rm d}M}(M_0) \: \Delta M_c}
\end{equation}
where $\Delta M_c \ll M_0$, then a step $\Delta S$ corresponding to a change in
mass $\Delta M$ larger than the mass resolution becomes a $2\sigma$ event. We
must choose this timestep carefully --- if it is too big, then the two
progenitor approximation will break down badly. If it is too small, then the
results become dominated by numerical noise. The above expression is
approximate, but provides a rule of thumb. Note that it scales with $M_0$
through the differential ${\rm d}S/{\rm d}M(M_0)$, so larger parent halos will
require smaller time steps.

In the simplest version of this algorithm, we start from a parent halo with
mass $M_0$ at $z_0$ and obtain the timestep $\Delta \omega$ from
Eqn.~\ref{eqn:timestep}. We work backwards in time from this point. We
choose a Gaussian random variable with unit variance and translate this to a
step $\Delta S$ using the transformation mentioned above. The new halo mass at
the earlier time $t(\omega + \Delta \omega)$ is then $M(S + \Delta S)$. LC93
argue that for a small enough timestep, all mergers may be treated as
binary. This makes the process very simple --- at each stage we break the halo
into two pieces with mass $M$ and $\Delta M \equiv M_0-M$ where $M$ is chosen
from the probability function $f(M)$. If the progenitor obtained in this way
is larger than $M_l$, we treat it as the next parent and repeat the
procedure. If it is smaller than $M_l$, then we treat it as accreted mass and
do not follow its history. This is essentially the same algorithm proposed in
LC93 at the top of page 641, and is similar to a generalized version of the
block model of \citeN{block1} and \citeN{block2}.

\begin{figure}
\centerline{\psfig{file=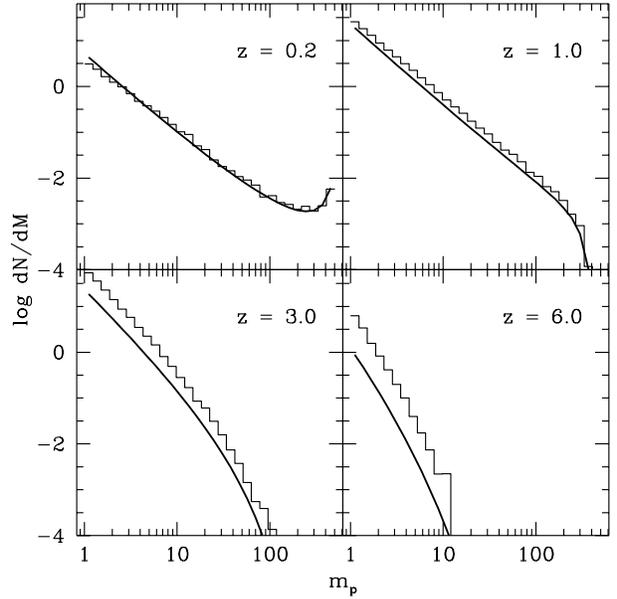,width=\columnwidth}}
\caption{The number of progenitors with mass $M$ for a halo with initial mass
$M_0 = 500$, at various redshifts as shown on the figure. The solid lines are
the predictions of the extended Press-Schechter theory. The histograms show the
results for merger trees constructed using the Binary Tree Method without
accretion, which is essentially the algorithm proposed by
\protect\citeN{lc:93}. The trees have an excess of halos compared to the
Press-Schechter model, and the discrepancy increases with redshift.}
\label{fig:bin_no_acc}
\end{figure}
\begin{figure}
\centerline{\psfig{file=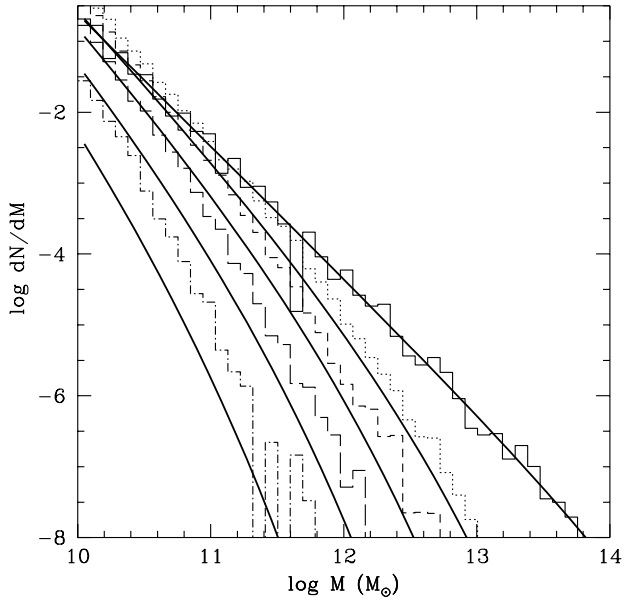,width=\columnwidth}}
\caption{The total mass function obtained from the Binary Tree Method without
accretion. All mergers are assumed to involve exactly two halos. The total mass
function is obtained by combining a grid of halos from $1\: M_l$ to $5\times
10^4 \: M_l$ ($1.0\times 10^{10} \msun$ to $5.0\times 10^{15} \msun$) weighted
with the Press-Schechter number density at $z=0$. The bold line is the
prediction of the Press-Schechter theory, at $z=0.16$, 2.5, 3.6, 5 and 7 from
top to bottom. The merger trees (histograms) overpredict the number of halos by
more than an order of magnitude after many steps in redshift. }
\label{fig:mf_bin}
\end{figure}

This approach has several advantages. It is simple and may be coded recursively
in a few lines. Because it mainly involves picking Gaussian random deviates, it
is also very fast. Rather than being imposed on an artificial grid in redshift
like previous methods, it reflects the intrinsic merging timescales of halos of
different mass contained in the extended Press-Schechter theory. Unfortunately,
the mass function of halos obtained in this way begins to develop an excess of
halos compared to the extended Press-Schechter model. This excess becomes more
and more severe as the number of steps increases. We show this for the
conditional mass function for a halo with an initial mass $M_0= 500 M_l$ in
Figure~\ref{fig:bin_no_acc}. This problem becomes quite serious when we combine
the merger histories of a grid of halos, with the appropriate Press-Schechter
weighting for the parent at $z=0$, to obtain the total mass function, shown in
Figure~\ref{fig:mf_bin}. The mass function reconstructed in this way should
yield exact agreement with the original Press-Schechter expression for the
universal mass function. With this method, the number of halos is overpredicted
by almost an order of magnitude by a redshift of $z\sim6$. This is especially
troublesome as the Press-Schechter model already gives an $\sim$30-50\% excess
of small halos compared to N-body simulations \cite{lc:94,gross:suites}. The
problem that we demonstrate here may explain why LC93 and \citeN{lc:94} find
that a similar Monte Carlo method leads to halo formation times that are 40\%
higher than the analytic predictions or the N-body results.

We believe that this problem is due to the simplifying assumption of binary
mergers. Because the merger rate of very small halos becomes effectively
infinite for CDM-like power spectra, non-binary mergers, at least involving
small halos, cannot be neglected. This statement is complementary to our
original premise regarding the importance of what we have called accreted
mass. One might think that we have simply not chosen a small enough time step,
but if this were the case the results should improve steadily as we decrease
the time step. We do not observe this behavior even for extreme reductions in
the time step.

Another way of stating the problem is that we have actually violated item 2 in
our list in \S\ref{sec:difficulties}. The first mass is chosen from the
distribution $f(M)$, but the mass of the second progenitor is not, it is just
assumed to be whatever mass is left over. This means that for every progenitor
with mass $M$ we always get a progenitor with mass $M_0-M$. It is easy to see
that this will lead to inconsistencies with the mean distribution function
$P(M)$.

\subsection{Binary Trees with Accreted Mass}
\label{sec:binacctrees}
We now attempt to cure the problem noted above by relaxing the simplifying
assumption of binary mergers from the previous subsection. Namely, we postulate
that mergers can involve at most two \emph{progenitors} (halos with $M> M_l$),
but an arbitrary number of halos with mass less than $M_l$. This of course
amounts to allowing for accreted mass. At any branching we may have only
accreted mass (zero progenitors), or alternatively one or two progenitors plus
accreted mass. In addition, the progenitor masses must always be picked from
the probability distribution $f(M)$. Leftover mass can contribute to accretion
but cannot be used for progenitors.

\begin{figure*}
\centerline{\psfig{file=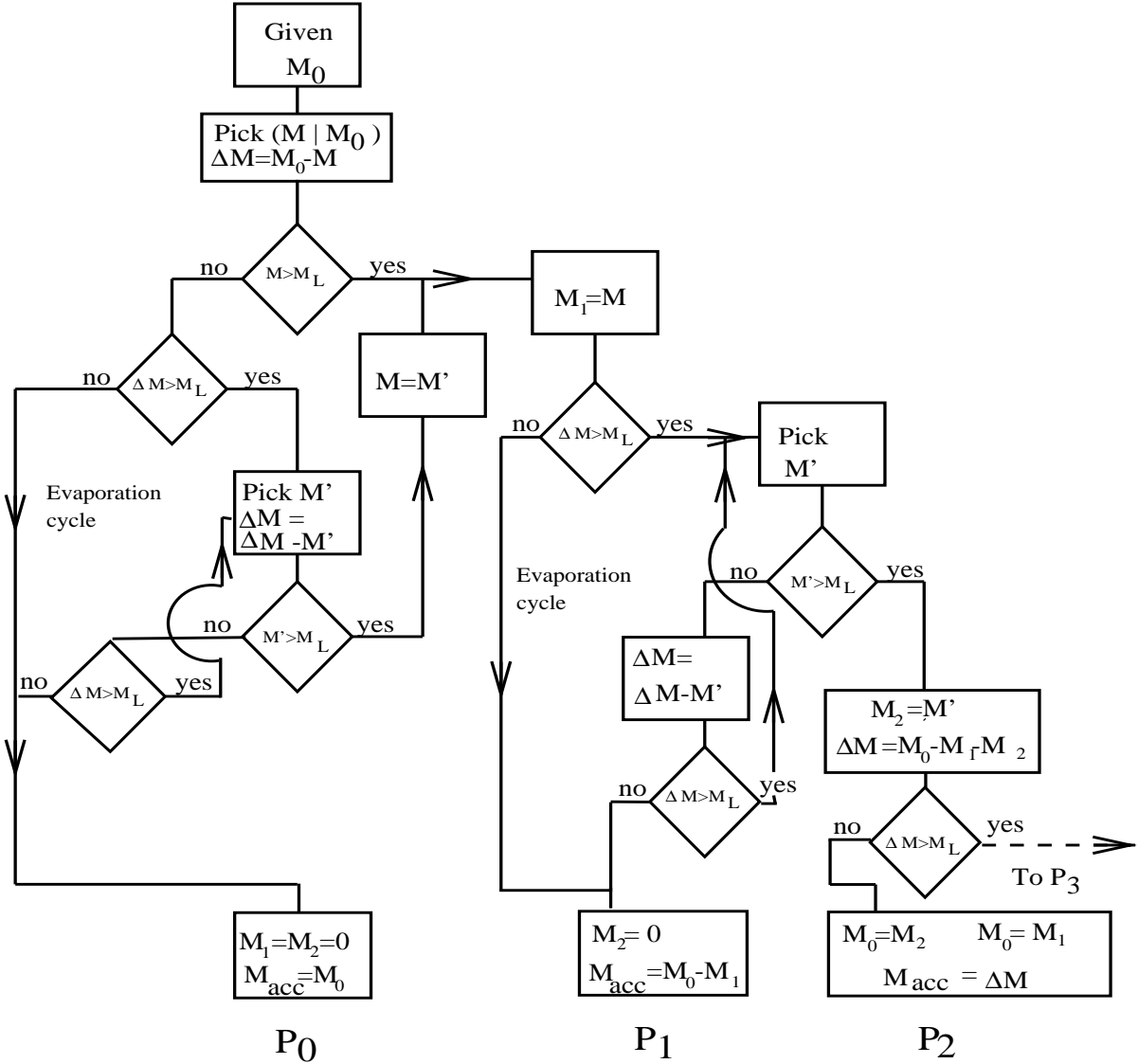,height=15cm,width=16cm}}
\caption{A flow-chart for one redshift step of the disintegration of a halo in
a merging tree. The end points ($\pr_1, \pr_2$, etc.) lead to identical
flow-charts for the subsequent timestep, with a new parent mass $M_0$. A
detailed discussion of the algorithm and this flow-chart is given in
\protect\S\ref{sec:binacctrees}.}
\label{fig:flow_chart}
\end{figure*}
The new recipe is as follows. The algorithm is shown in flow-chart form in
Figure~\ref{fig:flow_chart}. Given the parent mass $M_0$ we compute the time
step $\Delta \omega$ as before. Using this timestep throughout the following
steps, we:
\begin{enumerate}
\item
Pick a mass $M$ from the mass-weighted probability distribution
Eqn.~\ref{eqn:prob}. This mass can be anywhere in the range $0 \leq
M \leq M_0$. If $M < M_l$, we count it as accreted mass. If $M \geq
M_l$, we count it as a progenitor.
\item
Compute the unallocated mass $\Delta M = M_0-M$. 
\item
If the unallocated mass $\Delta M$ is larger than $M_l$, then it may or
may not contain a progenitor. To determine this, pick another mass $M$
from the distribution, but with the restriction $M<\Delta
M$. Depending on its mass, count it as accreted mass or a progenitor
as before. In either case, subtract $M$ from the mass reservoir.
\item
Repeat this process until either 
\begin{itemize}
\item
The mass reservoir $\Delta M$ falls below the minimum halo mass $M_l$,
in which case it must abandon any aspirations of harboring a real
progenitor and must be accreted mass, 
\item 
OR we have found a total of two progenitors ($M > M_l$), in which case
the remaining mass is considered to be accreted mass in accord with
our ansatz. 
\end{itemize}
\item
Each progenitor now becomes a parent, we calculate a new time step,
and repeat the whole process.
\end{enumerate}
In the flow-chart, branches leading to the outcome of zero, one and two
progenitors are labeled $\pr_0$, $\pr_1$, and $\pr_2$, in connection with the
formalism developed in the previous section.

Note that this procedure does not strictly fulfill requirement $2$ of equal
treatment of progenitors regardless of the order in which they are picked. This
inequality is necessary due to the mass conservation requirement.

\begin{figure}
\centerline{\psfig{file=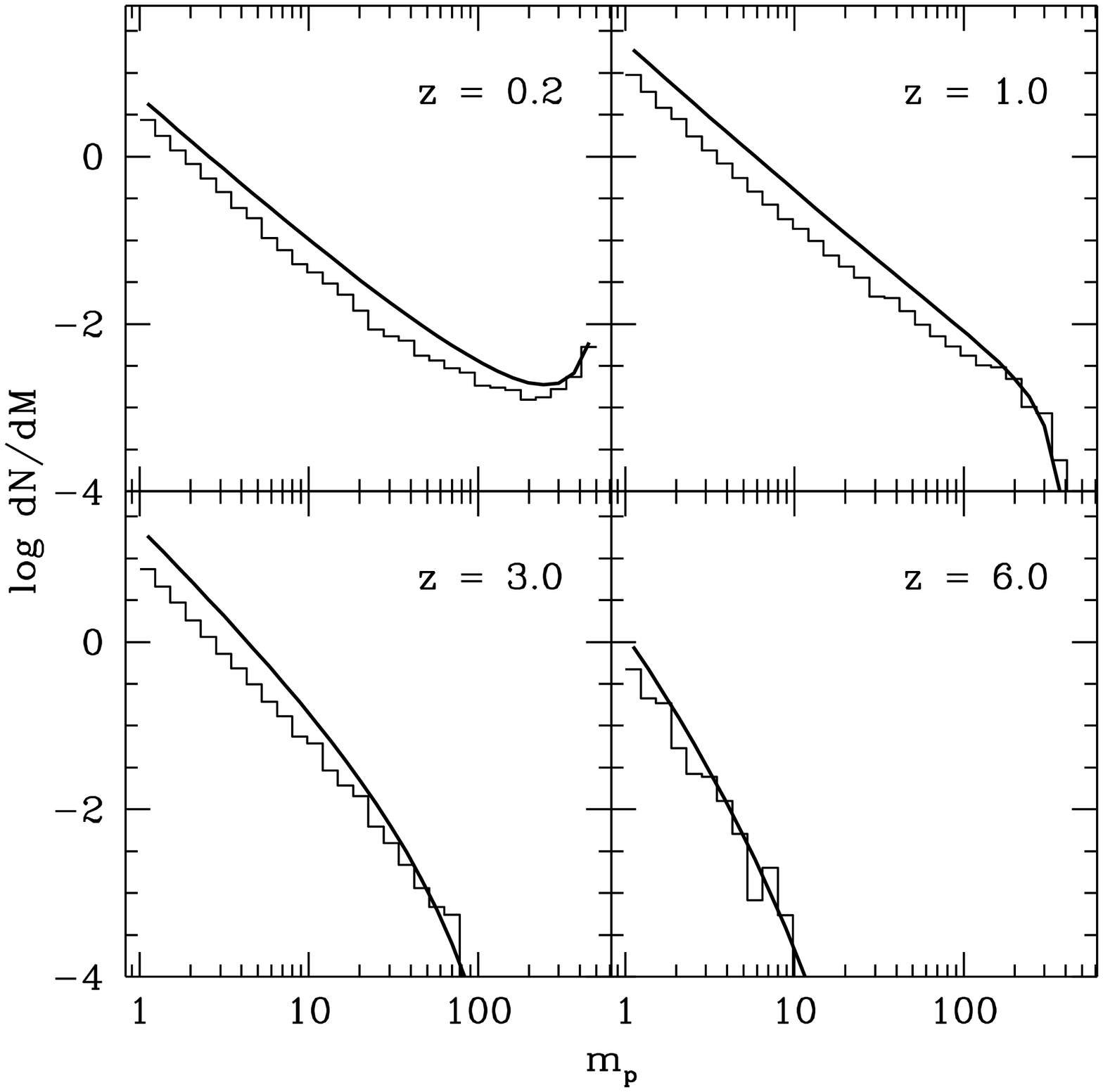,width=\columnwidth}}
\caption{The number of progenitors with mass $M$, for a halo with $M_0=500$,
using the Binary Tree Method with accretion. Here mergers may involve only two
halos with mass greater than $M_l$, but an indefinite number of mergers with
halos with mass smaller than $M_l$ (accretion). As usual the solid line is the
extended Press-Schechter prediction. The trees (histograms) now
\emph{underproduce} halos compared to the Press-Schechter model.}
\label{fig:bintree+acc}
\end{figure}
The results for the conditional mass function of a single halo are
shown in Figure~\ref{fig:bintree+acc}. The discrepancy is now in the
opposite direction: the number of halos is underpredicted relative to
the extended Press-Schechter prediction. Apparently this procedure
now overestimates the accreted mass. This is not too surprising since
we allowed large amounts of mass to be designated as accreted mass
simply because two progenitors had already been found. It appears that
even in the limit of small time steps and for large mass halos,
mergers between more than two halos cannot be neglected. In some ways
this is not surprising either, because after all the division into
$M>M_l$ and $M< M_l$ is arbitrary and has no physical basis.

\subsection{The Successful Method: $N$-Branch Trees With Accretion}
\label{sec:final}
It is trivial to generalize our previous recipe to allow an unrestricted number
of progenitors. We now continue picking progenitor masses until the unallocated
mass $\Delta M$ is less than $M_l$. This is indicated on the flow-chart by the
dashed line labeled ``To $\pr_3$''. Note that the total accreted mass can still
exceed $M_l$ because some of the attempts to pick progenitors yield halos with
$M < M_l$ and contribute to the accreted mass. We still pick the time-step so
that the number of progenitors cannot get \emph{too} large (we find that we
never exceed ten progenitors per time step even for cluster mass halos ($M_0 =
5 \times 10^{4}\, M_l$)). We find a good compromise between efficiency and
accuracy if we introduce an additional scaling in the expression for $\Delta
\omega$ from Eqn.~\ref{eqn:timestep}, of the form $b + a\log_{10} (M/M_l)$,
where we have used the parameters $a=0.3$ and $b=0.8$ for the results shown
here. This optimal scaling would change for a different power spectrum shape.

\begin{figure}
\centerline{\psfig{file=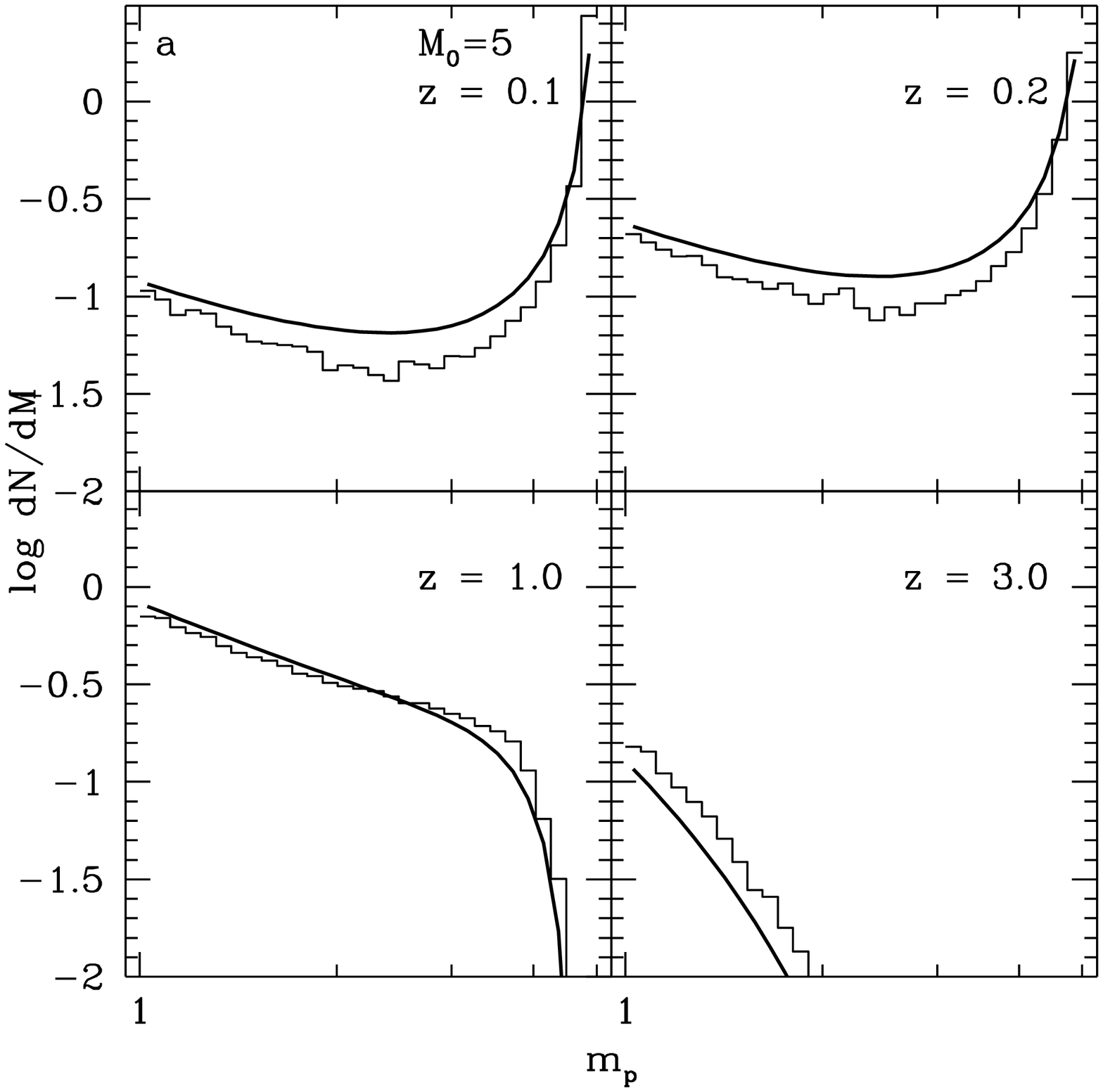,width=7cm}}
\centerline{\psfig{file=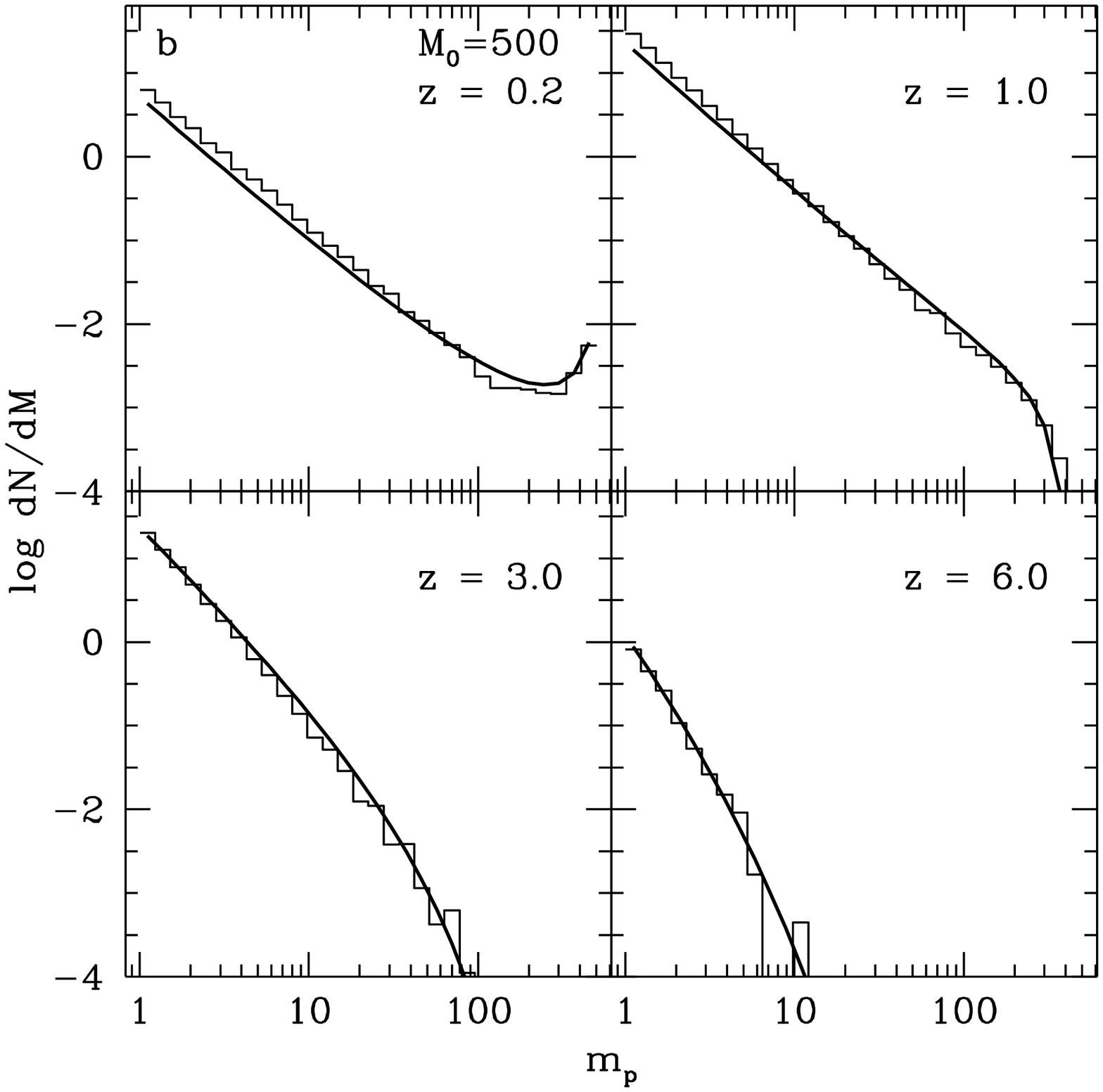,width=7cm}}
\centerline{\psfig{file=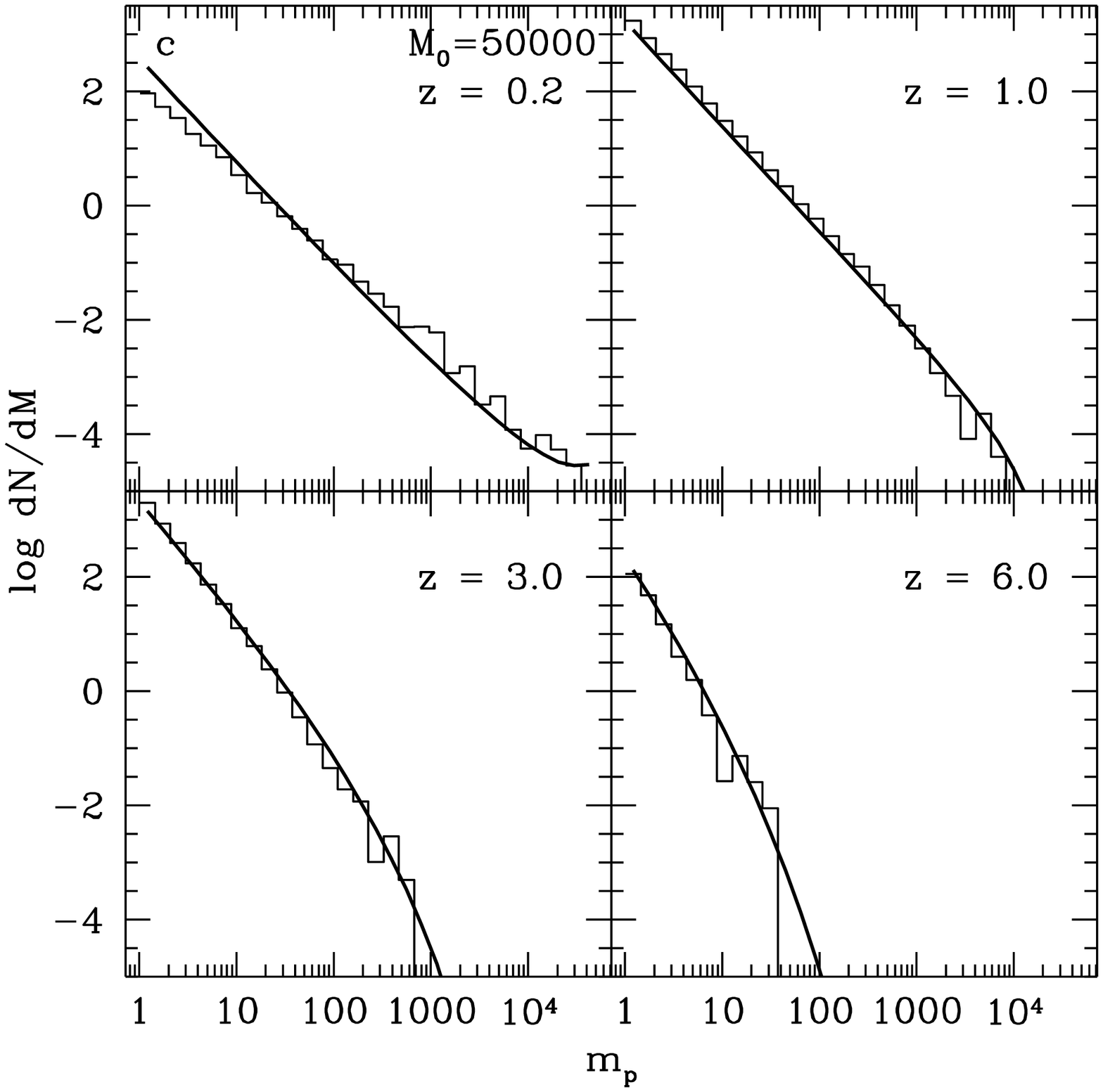,width=7cm}}
\caption{The number of progenitors with mass $M$, for a halo with initial mass
$M_0$, using the $N$-Branch Tree Method with accretion. This is the same as the
binary tree method with accretion except that an arbitrary number of
progenitors is allowed at each branching. The solid line is the extended
Press-Schechter prediction. (a)~$M_0=5\: M_l$ (b)~$M_0=500\: M_l$
(c)~$M_0=5\times 10^4\: M_l$. The merger trees (histograms) are in reasonably
good agreement with the extended Press-Schechter model. }
\label{fig:ntree}
\end{figure}
\begin{figure}
\centerline{\psfig{file=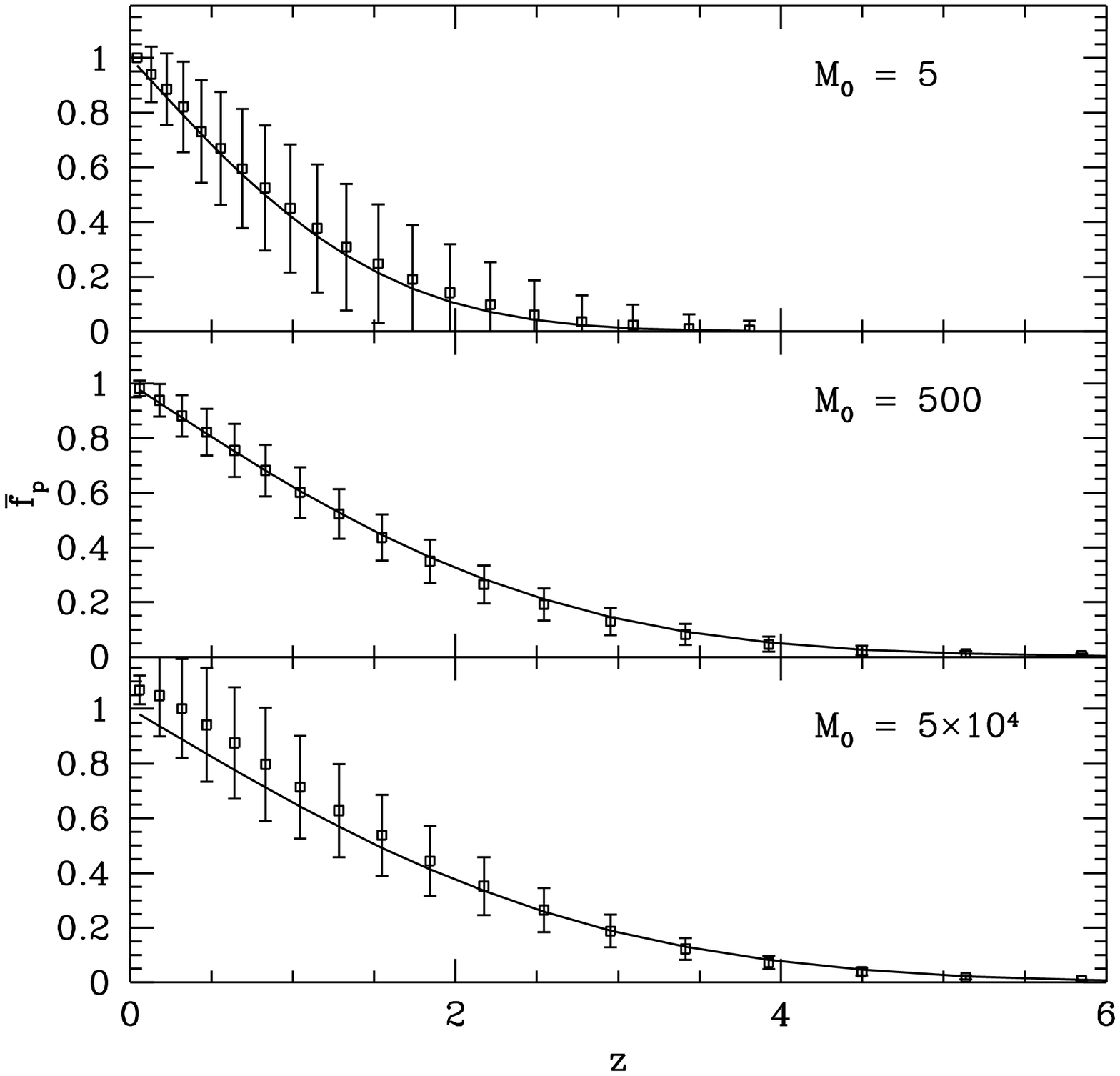,width=\columnwidth}}
\caption{The average fraction of the original mass $M_0$ contained in
progenitors (halos with $M> M_l$) at a redshift $z$. The solid line shows the
prediction of the Press-Schechter model. The square symbols show the average
given by the merger trees ($N$-Branch Tree Method with accretion) Error bars
show the standard deviation ($1\sigma$) over many ensembles.}
\label{fig:fp}
\end{figure}
\begin{figure}
\centerline{\psfig{file=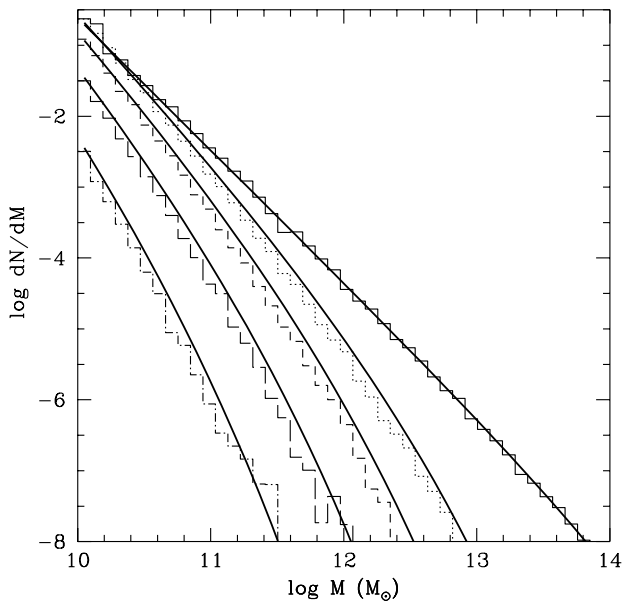,width=\columnwidth}}
\caption{The total mass function obtained from the $N$-Branch Tree Method with
accretion (the successful method). The solid lines show the prediction of the
Press-Schechter theory, at $z=0.16$, 2.5, 3.6, 5 and 7 from top to
bottom. Broken histograms show the mass function from the merger trees,
constructed as described in Fig.~\protect\ref{fig:mf_bin}. }
\label{fig:mf}
\end{figure}

This recipe gives good results for the conditional mass function for parent
halos with a wide range of masses. We show this in Figure~\ref{fig:ntree} for
parent halos with $M_0=5\, M_l, 500\, M_l$ and $5 \times 10^4 \, M_l$. The
agreement is poorest for halos with $M_0 \la 10\, M_l$. If we require strict
mass conservation, this is an unavoidable problem due to the shape of the
conditional mass function for halos of this size. This should be kept in mind
when setting the value of $M_l$ --- it should be chosen such that only objects
larger than $\sim 10M_l$ correspond to observable galaxies. We also check the
mass weighted quantity $\bar{f}_p$, the fraction of mass in progenitors as a
function of redshift. This is shown in Figure~\ref{fig:fp}.
This quantity shows good agreement for the smallest halo,
$M_0=5\, M_l$, which shows that the accreted mass is being treated properly, so
that we do not need to worry about the less than perfect agreement in the
conditional mass function, as long as the condition on $M_l$ mentioned above is
satisfied. Note that the agreement of the mass function can be improved by
adjusting the time-step, but at the expense of $\bar{f}_p$. We adjust the
time-step to achieve the best possible agreement for both the number and mass
weighted quantities, over the entire mass range. We compute the universal mass
function from the weighted grid of merging histories constructed using our new
scheme, and plot this in Figure~\ref{fig:mf}. We now find very good agreement
with the prediction of the standard Press-Schechter theory.

We therefore conclude that although this method is not rigorous, it produces
acceptable agreement with the mean quantities that we can check with the
Press-Schechter model.

\section{The Number-Mass Distribution of Progenitors}
\label{sec:distribution}
We have now developed a convenient and efficient method for constructing merger
trees. The averages derived from an ensemble of these trees agree with the
important mean quantities predicted by the extended Press-Schechter
theory. However, part of the motivation for developing this new method was to
ensure that the ensemble obeys the true joint probability distribution. We have
not yet shown this to be the case, and we have already mentioned the lack of
any information from the standard Press-Schechter formalism which would allow
us to evaluate the Monte Carlo trees against analytic predictions.

\begin{figure}
\centerline{\psfig{file=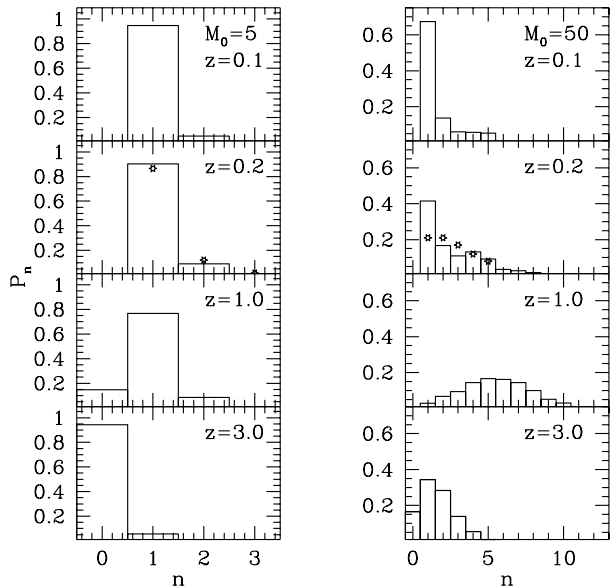,width=\columnwidth}}
\caption{The probability distribution of the number of progenitors, for a
parent halo with mass $M_0$ at several redshifts. The histogram shows the
results from the Monte Carlo merging trees (N-Branch Tree Method with
accretion). The stars show the results of the semi-analytic model developed in
section~\ref{sec:gestalt}. Left panel: $M_0=5\: M_l$. Right Panel: $M_0=50\:
M_l$}
\label{fig:pn}
\end{figure}
We do have the predictions of the model we developed in
section~\ref{sec:gestalt}, but we do not trust them for reasons discussed in
that section. However, out of curiosity we compare the predictions of the
semi-analytic accretion model for the probability distribution of the number of
progenitors, $\pr_N$, with the results of the Monte Carlo merger trees. The
integrals were performed numerically using a recursive, adaptive stepsize
Runga-Kutta algorithm. Figure~\ref{fig:pn} shows the $\pr_N$ distributions for
halos with mass $M_0=5$ and $M_0=50$ at redshifts of 0.1, 0.2, 1, and 3. The
results agree rather well for redshift steps of $\Delta z=0.2$ or smaller. For
larger steps in redshift, the semi-analytic results do not agree as well,
probably due to the breakdown of the accretion model.

\begin{figure}
\caption{The evolution with redshift of the two-dimensional distribution of the
number of progenitors and their masses for a halo of mass $M_0$, obtained from
Monte Carlo realizations of the $N$-Branch Merger Trees with
accretion. (a)~$M_0=5\: M_l$ (b)~$M_0=50\: M_l$ (c)~$M_0=500\: M_l$. 
These figures are available by anonymous ftp from ftp.ucolick.org
(pub/outgoing/rachel). 
}
\label{fig:pnm}
\end{figure}
Figure \ref{fig:pnm} demonstrates the importance of taking into account the
joint distribution in the progenitor number -- progenitor mass space. The
figure shows strong correlations between the two variables. Some of the
correlations are obvious: in Figure \ref{fig:pnm}a, we show the distribution
for a parent halo of $M_0=5\,M_l$. In the first redshift step ($z_0=0$,
$z_1=0.2$) the highest probability is obtained for having a single progenitor,
and this progenitor naturally contains a large fraction of the mass $M_0$. As
we progress in redshift, this correspondence is not preserved. Accreted mass
starts to be more and more significant, and the unseen accreted mass
complements low mass progenitors so that the sum may reach $M_0$. At the
formation epoch of $M_0$, all $\pr_N$ are populated, and in earlier stages most
of the halos go below $M_l$, when the dominant process is of single progenitors
that accumulate accreted mass. It is interesting to notice that the ``formation
epoch'' (the earliest time when the largest progenitor has mass greater than
$M_0/2$) is not dominated by mergers of equal mass halos, but rather approached
via slow accretion. This general picture also remains valid for higher $M_0$
(Figure~\ref{fig:pnm}b and c): an increase in the number of progenitors occurs
towards an intermediate redshift, and it then declines towards containing most
of the mass in the accreted component. However it should be noted that the
highest mass considered here would be comparable to the halo of a Milky Way
sized galaxy ($5 \times 10^{12} \msun$). For much larger mass halos (comparable
to group or cluster mass halos), accretion is less important relative to the
aggregation of roughly equal mass progenitors.

The probability for the number of progenitors spans substantial parts of its
permitted range even for $M_0=500$. It is therefore clear why an infinitesimal
step is needed for the two-progenitor scheme to work. As soon as we consider a
finite timestep, the probability for $\pr_{N>2}$ is no longer negligible.

The details of the redshift sequence represent the characteristics of the
specific power spectrum and cosmology we used for the Monte-Carlo
realization. The formation time as a function of halo mass and redshift are
determined by the cosmology and the power spectrum. The qualitative trend of
this sequence, however, is similar for all cosmologies and all hierarchical
power spectra.

Figure \ref{fig:pnm} suggests that the progenitor mass and number distribution
functions are an interesting avenue to pursue in the study of structure
formation via merger trees. More importantly, it points at the existing
interplay between the accreted mass, the progenitor mass and the number of
progenitors; an interplay in which none of the three can be treated separately
from the other.

\section{Summary and Conclusions}
\label{sec:sum}
We have presented a new method for constructing the merging history of dark
matter halos in a semi-analytic way. We have highlighted the need to impose an
arbitrary mass cutoff for practical reasons, which leads us to distinguish
between halos above and below this threshold as ``progenitors'' or ``accreted
mass'', respectively. The scheme we have proposed and implemented treats
accreted mass and progenitors in a self consistent way, and produces good 
agreement with the average quantities predicted by the underlying
Press-Schechter theory, such as the conditional and universal mass function of
halos and the mean mass in progenitors as a function of redshift.

Our method is an improvement on the method proposed by \citeN{lc:93}, which,
after many steps in redshift, substantially overproduces halos relative to the
Press-Schechter mass function. Our work suggests that it is not possible to
simultaneously conserve mass exactly and retain the exact agreement with the
conditional mass function from the extended Press-Schechter model. The method
of \citeN{kw} reproduces the conditional mass function exactly and conserves
mass approximately. Our method conserves mass exactly and reproduces the
conditional mass function approximately. This seems to be a necessary
trade-off. Our method does have certain practical advantages: it does not
require the creation and storage of a large number of ensembles, it is
numerically robust, it does not require the imposition of a grid is mass or
redshift, and it will work for any power spectrum.

We have pointed out the necessity of investigating the full probability
distribution of the number of progenitors and their masses. This cannot be
tested within the boundaries of the existing theory, and so must be examined by
comparisons with N-body simulations. However, the simulations have their own
problems and complications, such as the limitations of mass and spatial
resolution and the ambiguities of defining halos, so they should not be
regarded as necessarily representing the absolute truth. In addition the
agreement between the simulations and the Press-Schechter model is only
approximate, even for the mean quantities such as the mass function. It would
therefore be desirable to have a reliable theoretical means of addressing this
problem. We have attempted to reformulate the extended Press-Schechter theory
to obtain the full probability distribution for the number of progenitor halos
$\pr_N$. Although this model gives qualitatively reasonable results for certain
cases, some ingredients remain ad-hoc.

For the moment, this leaves us with no recourse but to appeal to N-body
simulations. In a companion paper \cite{paperII}, we will compare the results
we have obtained here with numerical simulations. This comparison has two
goals: {\bf (a)} to determine the quality of agreement of the analytic and
Monte Carlo results with the N-body simulation results. {\bf (b)} to study the
full distribution of the various quantities, and determine whether the Monte
Carlo method developed here reproduces these results. The merger trees will
then be used as the framework for the development of full semi-analytic galaxy
formation models, and used to compare with a variety of observations
\cite{mythesis,sp:98}.

\section*{Acknowledgements}
We would like to thank Gerard Lemson for useful discussions and healthy
skepticism. We also thank Guinevere Kauffmann and Simon White for their
thorough reading of the manuscript, an important demonstration of some
drawbacks in the proposed procedure, and comprehensive discussions. We thank
Sandy Faber and Joel Primack for detailed comments on the manuscript and useful
insights. RSS acknowledges support from an NSF GAANN fellowship. This work was
supported in part by NASA ATP grant NAG5-3061.

\bibliographystyle{mnras}
\bibliography{mnrasmnemonic,merge}

\end{document}